\documentclass[aps,prl,twocolumn,groupedaddress,showpacs]{revtex4}
\usepackage{amsmath,amssymb}
\usepackage[dvips]{graphicx}

\begin{document}
\title{Complex solitons with power law behaviour in Bose-Einstein condensates near Feshbach resonance}
\author{Utpal Roy$^{\mathrm{1}}$}

\author{Rajneesh Atre$^{\mathrm{1,2}}$}

\author{C. Sudheesh$^{\mathrm{1}}$}

\author{C. Nagaraja Kumar$^{\mathrm{3}}$}

\author{Prasanta K. Panigrahi$^{\mathrm{1}}$}
\email{prasanta@prl.res.in}
\affiliation{$^{\mathrm{1}}$Theoretical Physics Division, Physical Research Laboratory,
Ahmedabad 380 009, India}
\affiliation{$^{\mathrm{2}}$School of Physics, University of
Hyderabad, Hyderabad 500 046, India}
\affiliation{$^{\mathrm{3}}$Department of Physics, University of
Panjab, Chandigarh 160 014, India}

\date{\today}
\begin{abstract}
Complex, localized stable solitons, characterized by a power law
behaviour, are found for a quasi-one-dimensional Bose-Einstein
condensate near Feshbach resonance. Both dark and bright solitons
can be excited in the experimentally allowed parameter domain, when
two and three-body interactions are respectively repulsive and
attractive. These solutions are obtained for non-zero chemical
potential, unlike their unstable real counterparts which exist in
the limit of vanishing $\mu$. The dark solitons travel with constant
speed, which is quite different from the Lieb mode, where profiles
with different speeds, bounded above by sound velocity can exist for
specified interaction strengths.
\end{abstract}
\pacs{03.75.Kk,03.75.Lm,05.45.Yv} \maketitle

The dynamics of non-linear waves in Bose-Einstein condensate (BEC)
is a subject of immense theoretical and experimental interest in
current literature. The recent observation of dark \cite{burger},
bright solitons \cite{khawaja,khaykovich,strecker1,wieman1}, soliton
trains \cite{strecker} and Faraday waves \cite{staliunas} have given
considerable impetuous to the investigation of the formation
mechanism and control of various non-linear excitations in the
quasi-1D scenario \cite{atre}. The mean field equation, governing
the dynamics of BEC, is the non-linear Schr\"{o}dinger equation with
a harmonic trap. The non-linearity originating from the two-body
interaction is characterized by the s-wave scattering length $`a$',
which can be controlled through Feshbach resonance \cite{Inouye}, as
also the width of the transverse profile \cite{staliunas}. For
$a>0$, elastic interaction is repulsive and the BEC is stable.
Negative scattering length implies attractive interaction, where the
condensate is found to be stable up-to a certain limit of the number
of atoms \cite{dalfovo,dodd,hulet}.

The three-body interaction can be generally treated as a
perturbation over the two-body case; it becomes significant for
short range and larger scattering length, as is the case near
Feshbach resonance. For a dense atomic media also, three-body
interaction plays an important role. It is estimated for
$\rm{Rb}$-BEC that, the real part of the three-body interaction term
is $10^3-10^4$ times larger than the imaginary part
\cite{aurel,leanhardt,pieri,koh}. Hence, we do not consider the
three-body recombination here, when the corresponding coupling
constant is imaginary. A number of theoretical studies have been
carried out considering three body interaction in both three- and
quasi-one-dimensions
\cite{kolo1,akhmediev,kolo2,brazh,pardo,axzhang}. Localized soliton
solutions of both elliptic function and power-law type have also
been investigated \cite{bhaduri,ogren,khay}. Dark soliton of secant
hyperbolic form manifested in purely repulsive three-body
interaction regime \cite{ogren}, relevant for Tonks-Girardeau gas
\cite{tonks,paredes,kinoshita}. In this case, the soliton velocity
is bounded above by sound velocity. Real solitons of both types were
also analyzed in \cite{khay}, where the algebraic one was found to
exist only in the $\mu\rightarrow0$ limit and was unstable.

In this letter, we demonstrate the existence of power-law type
complex solitons in the presence of repulsive two- and attractive
three-body interactions. Unlike the real case, the obtained dark and
bright soliton solutions can exist for non-vanishing $\mu$ and are
stable. The dark solitons have a constant velocity determined by the
interaction strengths, which is quite different from the Lieb mode
case \cite{ogren}. Their profiles can change as a function of the
parameters of the theory. The corresponding velocities change from
zero to sound velocity. Interestingly, in the parameter domain where
soliton velocity equals sound velocity, it is found that the
Bogoliubov dispersion is of the quadratic type. For specificity, we
consider $^{87}\rm{Rb}$ with $m=1.44\times 10^{-25}$ Kg and the
axial density $\sigma_0$ in the range $5.43 \times
10^{7}\rm{cm}^{-1}-9.67\times 10^{8}\rm{cm}^{-1}$. The transverse
trap-frequency is taken as $\omega_{\perp}=2\pi\times 140$ rad/sec
and the two-body coupling constant $g_2=4.95 \hbar \times 10^{-11}
\rm{cm}^3 /\rm{sec}$. The three-body interaction coefficient $g_3$
has already been estimated
\cite{koh,hammer1,hammer2,hammer3,tewari}. The above parameters
allow the present solitons to exist, in the domain of $g_3$, taking
values from $-10^{-27} \rm{cm}^6 /\rm{sec}$ to $-10^{-26} \rm{cm}^6
/\rm{sec}$ (scaled by $\hbar$). This is in the range of
theoretically predicted value for $^{87}\rm{Rb}$.

A linear stability analysis using spectral method is carried out,
which shows that the obtained solutions are stable against small
perturbations in both dark and bright soliton regimes.
Modulational instability (MI) analysis
\cite{hasegawa,tai,GPAgarwal} reveals that the parameter regimes
relevant for the solutions are away from the domain of
instability.

\begin{figure}[htpb]
\begin{center}
\includegraphics[width=3.2 in]{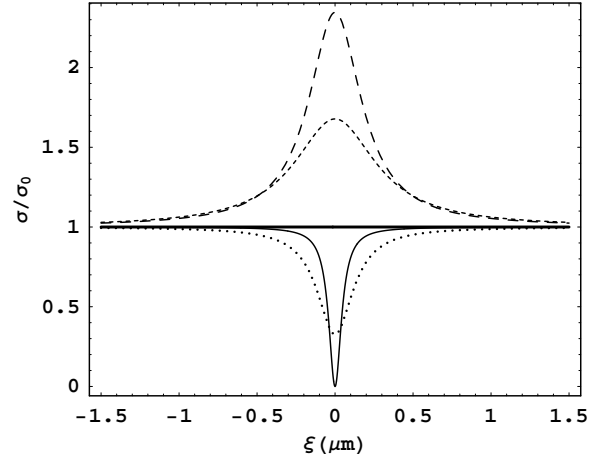}
\caption{ The density profiles of soliton solutions for different
three-body interactions with $g_2=4.95 \hbar\times 10^{-11}
\rm{cm}^3 /\rm{sec}$. The obtained dark solitions for $\tilde{g_3}=-
\tilde{g_2}/2\sigma_0$ (solid line), $\tilde{g_3}=-0.45
\tilde{g_2}/\sigma_0$ (dotted line) and bright solitons for
$\tilde{g_3}=- 0.32 \tilde{g_2}/\sigma_0$ (small-dashed line),
 $\tilde{g_3}=-0.28 \tilde{g_2}/\sigma_0$ (long-dashed line).
The thick  solid line represents the constant background density
$\sigma_0$ for $\tilde{g_3}=- 3\tilde{g_2}/8\sigma_0$.}
\label{profilefig}
\end{center}
\end{figure}

The 3D Gross-Pitaevskii (GP) equation for the wave function
$\Psi(r,t)$, with an additional three-body interaction, is given by
\begin{equation}
i\hbar \,\frac{\partial \Psi}{\partial t}=-\frac{\hbar^2}{2m}
\nabla^2 \Psi+\big(V+ g_2\,|\Psi|^2 + g_3|\Psi|^4
-\mu\big)\,\Psi,
\label{twothree}
\end{equation}
where $\mu$ is the chemical potential. The cylindrical harmonic trap
is given by $V=m\omega_{\perp}^{2}(x^2+y^2)$/2 with a tight
transverse confinement. For sufficiently small transverse dimension
of the cloud, the wave function can be written as
$\psi(r,t)=f(z,t)\,\phi_0$ with $\phi_0= \sqrt{\frac{1}{\pi
a_{\perp}^2}} \exp(-\frac{x^2+y^2} {2 a_{\perp}^2})$ and
$a_{\perp}=\sqrt{\hbar/(m \omega_{\perp})}$. The longitudinal
envelope function $f(z,t)$ obeys \cite{jackson2,salasnich},
\begin{equation}
i\hbar \,\frac{\partial f}{\partial t}=-\frac{\hbar^2}{2m}
\frac{\partial^2 f}{\partial z^2}+\big(\tilde{g_2}|f|^2 + \tilde{g_3}|f|^4-\mu)\,f,
\label{envelope}
\end{equation}
where the reduced interaction coefficients are
\begin{equation}
\tilde{g_2}=\frac{m \omega_{\perp}}{2 \pi \hbar}\, g_2, \quad
\tilde{g_3}=\frac{m^2 \omega_{\perp}^2}{3 \pi^2 \hbar^2}\, g_3.
\end{equation}
For the space-time independent solution, chemical potential can be
written in terms of the asymptotic density $\sigma_0$;
\begin{equation}
\mu=(\tilde{g_2}+\tilde{g_3} \sigma_0)\sigma_0. \label{mueqn}
\end{equation}
The superfluid velocity is obtained from the continuity equation:
\begin{equation}
v=u(1-\frac{\sigma_0}{\sigma}), \label{imaginaryeqn}
\end{equation}
where $v=\frac{\hbar}{m}\frac{\partial\theta}{\partial \xi}$ and
$f=\sqrt{\sigma(\xi)}\, e^{i \theta(\xi)}$. The hydrodynamic
equation for the density is then,
\begin{eqnarray}
-\frac{\hbar^2}{2m} (\sigma_z^2-2 \sigma \sigma_{zz})&=&4 \tilde{g_3} \sigma^4+
4 \tilde{g_2} \sigma^3\nonumber\\&-&(4 \mu+2 m u^2)\sigma^2+2 m u^2\sigma_0^2.
\label{realeqn}
\end{eqnarray}
A power law ansatz
\begin{equation}
\sigma(\xi)=\sigma_0\left(1-\frac{B}{1+D \xi^2}\right),
\label{solution}
\end{equation}
is found to solve Eq.~\ref{realeqn}, where $B$ and $D$ are given by,
\begin{eqnarray}
B=\frac{3 \tilde{g_2}  + 8\tilde{g_3} \sigma_0 }{2 \tilde{g_3} \sigma_0}
,\quad
D=-\frac{m}{\hbar^2}\,\frac{(3 \tilde{g_2} +8 \tilde{g_3}\sigma_0)^2}
{6\tilde{g_3}},\nonumber\\
\quad {\rm with}\quad
u=\pm\Big(\frac{\tilde{g_2}\sigma_0+2\tilde{g_3}\sigma_0^2}{m}\Big)^{\frac{1}{2}}.\quad
\label{soln2}
\end{eqnarray}
It is transparent that, non-singular solutions exist only when
$\tilde{g_3}$ is negative, i.e., attractive three-body interaction.
The value of $\tilde{g_2}$ should be positive from the reality of
the soliton velocity, implying repulsive two-body interaction. As
mentioned before, $u$ is a constant for given parameter values and
density. This situation is quite different from the Lieb-mode case,
where the soliton velocity can take different values, bounded above
by the sound velocity. The obtained solutions can be categorized
into three different classes depending on the values of
$\tilde{g_3}$ for a given $\tilde{g_2}$ and $\sigma_0$: (i) A dark
solition in the range $-\tilde{g_2}/2\sigma_0\leq\tilde{g_3}<-3
\tilde{g_2}/8\sigma_0$, (ii) a constant background for
 $\tilde{g_3}=-3 \tilde{g_2}/8\sigma_0$ and (iii) a bright soliton for
 $-3\tilde{g_2}/8\sigma_0\leq\tilde{g_3}<-0.28 \tilde{g_2}/\sigma_0$. In these regimes $\mu$
is a real positive quantity. For $\mu=0$, one only obtains a real
soliton \cite{khay}. Figure \ref{profilefig} shows the density
profiles of dark and bright solitions for different values of
$\tilde{g_3}$. Usually, repulsive interaction alone creates dark
soliton, whereas attractive one results in bright solitons in BEC.
As both types of forces are present in the present system, one gets
dark and bright solitons, depending on the values of the coupling
constants $\tilde{g_2}$ and $\tilde{g_3}$. In Fig.~\ref{profilefig},
$\tilde{g_3}$ is increased from dark to bright soliton for a
particular value of $\tilde{g_2}$. The density profile smoothly
transits from dark soliton to the bright one. Hence, larger the
value of three-body interaction, greater is the accumulation of
atoms in the condensate. Physically it amounts to increasing the
local density of atoms for going from dark to bright regime. This
leads to a depletion of atoms in the background. The solid line in
Fig.~\ref{profilefig} corresponds to $u=0$ case. Thick solid line is
the homogeneous background $\sigma=\sigma_0$, where $u=\pm
\frac{1}{2}\sqrt{g_2 \sigma_0/m}$.

\begin{figure}[htpb]
\begin{center}
\includegraphics[width=2.8 in]{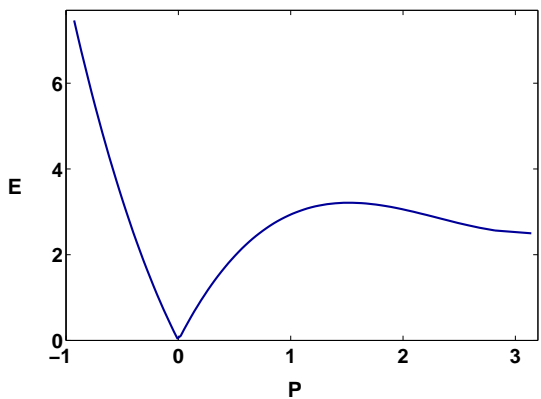}
\caption{Energy vs momentum for the dark and bright solitons for $-
\tilde{g_2}/2\sigma_0 \leq \tilde{g_3}\leq -0.32
\tilde{g_2}/\sigma_0$ with the same $g_2$ used in
Fig.~\ref{profilefig}. Energy and momentum are respectively scaled
by $\hbar^2 {\sigma_0}^2/m\times 10^{-4}$ and $\hbar \sigma_0$.}
\label{epplane}
\end{center}
\end{figure}

The appropriately normalized energy functional
\begin{eqnarray}
E=\int\Big[\frac{\hbar^2}{2\,m}&&\!\!\!\!\frac{\partial f}{\partial z}
\frac{\partial f^{*}}{\partial z}+\frac{\tilde{g_2}}{2}(f f^{*})^2-
\frac{\tilde{g_2}}{2}\sigma_0^2\nonumber\\
&&+\frac{\tilde{g_3}}{3}(f f^{*})^3-\frac{\tilde{g_3}}{3}
\sigma_0^3-\mu(ff^*-\sigma_0) \Big] dz,\nonumber
\end{eqnarray}
yields
\begin{equation}
E=\sqrt{\frac{3 \pi^2\hbar^2}{2\,m}}\,\,\frac{\tilde{g_2} \,| 3
\tilde{g_2}+8 \tilde{g_3}\sigma_0|}{16 |\tilde{g_3}|^{3/2}},
\end{equation}
for the soliton profile. Dark soliton with $u=0$ corresponds to
energy $E=\pi \hbar \sigma_0 \sqrt{3 \tilde{g_2} \sigma_0/(64 m)}$,
whereas it goes to zero when the background is uniform. Momentum of
the condensate profile
\begin{equation*}
P = \frac{-i\hbar}{2} \int dz [ f^* f_z -f^{*}_z f ] = m \int dz
(\sigma-\sigma_0) v(z),
\end{equation*}
gives
\begin{equation}
P=\pi\hbar\sigma_0\frac{u}{|u|}\Big(1-\sqrt{\frac{3(\tilde{g_2}+2\tilde{g_3}\sigma_0)}{-2\tilde{g_3}
\sigma_0}}\,\,\Big).
\end{equation}
It reaches maximum value ($P_{\rm max}=\pi\hbar\sigma_0)$ when
$\tilde{g_3}=-\tilde{g_2}/(2 \sigma_0)$. Figure \ref{epplane}
depicts the variation of energy with momentum for different
three-body interaction strengths. Positive momentum is the region of
dark soliton, where as negative one corresponds to bright soliton.
Energy and momentum vanish at the transition point
$\sigma=\sigma_0$. Dispersion graph is stiffer in the bright soliton
regimes.

The number of atoms in the condensate, normalized to vanish at
$\sigma=\sigma_0$,
\begin{eqnarray}
N&=&\int dz (\sigma_0-\sigma(z))\nonumber\\
&=&\left(\frac{3\pi^2\hbar^2 \sigma_0
^2}{m|\tilde{g_3}|}\right)^{1/2}|3\tilde{g_2}+8\tilde{g_3}\sigma_0|,
\end{eqnarray}
shows that the maximum deficiency of atoms in the dark soliton
regime is $N=(6\pi \hbar^2 \sigma_0 ^3 \tilde{g_2}/m)^{1/2}$.

We now analyze the dynamical stability of obtained solutions using
the spectral method \cite{su,bara}. A small perturbation
$e^{\lambda\,t} \phi(\xi)$ of soliton solution satisfies
\begin{equation}
\mathbf{A}\,\vec{\varphi}=\lambda\, \mathbf{J}\,\vec{\varphi},
\end{equation}
where $\vec{\varphi}$ is a two-dimensional vector and its components
are real and imaginary parts of the perturbation:
$\vec{\varphi}=(\phi_1\;\phi_2)^{T}$. Here, $\mathbf{J}$ is a
two-dimensional matrix with $J_{11}=J_{22}=0$ and
$J_{12}=-J_{21}=1$. The elements of the matrix operator $\mathbf{A}$
are
\begin{eqnarray}
A_{11}\!&=&\!\frac{\hbar^2}{2\,m}\frac{\partial^2 }{\partial \xi^2}\!-\!
\tilde{g_2}(3\,f_1^2+f_2^2)\!-\!\tilde{g_3}(5\,f_1^4+f_2^4+6\,f_1^2\,f_2^2)+\mu,\nonumber\\
A_{12}\!&=&\!\hbar\,u\frac{\partial}{\partial \xi}-
2\,\tilde{g_2}\,f_1\,f_2-4\,\tilde{g_3}\,f_1\,f_2\,|f|^2,\nonumber\\
A_{21}\!&=&-\!\hbar\,u\frac{\partial}{\partial \xi}-
2\,\tilde{g_2}\,f_1\,f_2-4\,\tilde{g_3}\,f_1\,f_2\,|f|^2\quad{\rm and}\nonumber\\
A_{22}\!&=&\!\frac{\hbar^2}{2\,m}\frac{\partial^2 }{\partial \xi^2}\!-\!
\tilde{g_2}(f_1^2+3\,f_2^2)\!-\!\tilde{g_3}(f_1^4+5\,f_2^4+6\,f_1^2\,f_2^2)+\mu,\nonumber
\end{eqnarray}
where $f=(f_1+ i\,f_2)$. The soliton solution is stable if real part
of the eigenvalue $\lambda$ is negative. $\phi_1$ and $\phi_2$ are
expanded into a spectral series over $800$ modes. This numerical
analysis shows that both bright and dark  solitons solutions are
stable in the entire domain of the solutions.

It is now worth investigating the issue of modulation instability
since the three-body interaction is attaractive. Phenomenon of
modulational instability has been extensively investigated in
literature for BEC \cite{salasnichMI,carr,solomonMI}. A single
component BEC with an attractive atom-atom interaction, can result
in modulational instability, when the density of atoms exceeds a
certain critical value. We assume $f=(f_0 +
\tilde{f})exp(i\tilde{\phi})$, where the infinitesimal fluctuation
$\tilde{f}$ is given by
\begin{equation}
\tilde{f}=\tilde{f_1} cos(K z-\Omega t)+i \tilde{f_2} sin(K z-\Omega
t).
\end{equation}
$\Omega$ and $K$ are respectively, the frequency and propagation
constant, of the modulated wave. The above transformation produces
two sets of equations involving $\tilde{f_1}$ and $\tilde{f_2}$. Non
trivial solutions are obtained only if $K$ and $\Omega$ satisfy the
dispersion relation, $2m\Omega^2= K^2 (\hbar^2 K^2/2m-4
|\tilde{g_3}| {f_0}^4 + 2 \tilde{g_2} {f_0}^2)$, where
$\tilde{g_3}<0$ and $\tilde{g_2}>0$. If $ \hbar^2 K^2/2m < (4
|\tilde{g_3}| {f_0}^4 - 2 \tilde{g_2} {f_0}^2 )$, it would show
modulation instability. This condition immediately implies $
\tilde{g_2} < 2|\tilde{g_3}| \sigma_0$, which is not in the allowed
parameter range for the obtained solutions. Thus our solutions are
modulationally stable.

In conclusion, complex soliton solutions with power law decay have
been identified in the quasi-one-dimensional GP equation with
repulsive two- and attractive three-body interactions. These
solutions, when superfluid velocity depends on density, show a
slower asymptotic decay compared to the elliptic function type
solutions. We have considered the parameters relevant to
$^{87}\rm{Rb}$, with the three-body coupling constant in the
theoretically predicted range. This opens the possibility of
observing these complex solitons in realistic BEC. Soliton velocity
is fixed by the strength of the interactions and are stable against
small perturbations. They are also modulationally stable. One would
like to study their behaviour in a trap for the purpose of coherent
control. The analysis of two-soliton sector is also an interesting
problem, as is the investigation in higher dimensions.

\end{document}